# Electrical-driven Plasmon Source on Silicon based on Quantum Tunneling

Hasan Göktaş[1,2], Fikri Serdar Gökhan[3], Volker J. Sorger[1*]

[1]Department of Electrical and Computer Engineering, George Washington University, Washington, D.C. 20052, USA
[2]Department of Electrical and Electronic Engineering, Harran University, Sanliurfa, 63000, Turkey
[3] Department of Electrical and Electronic Engineering, Alanya Alaaddin Keykubat University, Kestel, Alanya, Antalya, Turkey
[*]Email: sorger@gwu.edu

**An efficient silicon-based light source presents an unreached goal in the field of photonics, due to Silicon's indirect electronic band structure preventing direct carrier recombination and subsequent photon emission. Here we utilize inelastically tunneling electrons to demonstrate an electrically-driven light emitting silicon-based tunnel junction operating at room temperature. We show that such a junction is a source for plasmons driven by the electrical tunnel current. We find that the emission spectrum is not given by the quantum condition where the emission frequency would be proportional to the applied voltage, but the spectrum is determined by the spectral overlap between the energy-dependent tunnel current and the modal dispersion of the plasmon. Experimentally we find the highest light outcoupling efficiency corresponding to the skin-depth of the metallic contact of this metal-insulator-semiconductor junction. Distinct from LEDs, the temporal response of this tunnel source is not governed by nanosecond carrier lifetimes known to semiconductors, but rather by the tunnel event itself and Heisenberg's uncertainty principle.**

With the emergence of photonic integration [1] the challenges a) to create light from silicon and b) to realize electronically-compact photonics have delayed the anticipated introduction of photonics into electronic consumer products [2]. While the diffraction limit of light has been surpassed using polaritonic modes [3] demonstrating device functionality such as modulation [4], light emission [5,6], detection [7], and tunable metasurfaces [8], the search for an electrically-driven silicon-compatible light source operating at or above room temperature is yet outstanding. While Silicon is capable of light emission, explored devices and emission mechanisms are challenged by operation instability and efficiency [9]. As a result, the light source either requires hetero-material integration, flip-chip bonding, or must be considered off-chip altogether [10-13]. Other source-related challenges include temporal effects of the gain medium; with the spontaneous emission lifetime of semiconductors being about nanoseconds, any LED is limited to modulation rates of a ~GHz. The time response of a laser on the other hand is limited by the gain relaxation oscillations, but is still limited by optical non-linear gain-compression (gain saturation) effects to the GHz-range [5]. While the Purcell effect accelerates the spontaneous emission process [14] raising the modulation frequency of a LED [15], it also reduces the laser threshold via improving the pump efficiency by increasing the spontaneous emission factor, $\beta$ [16, 17]. Yet plasmon lasers require an inherently higher threshold power than photonic counterparts due to the lossy metals involved [18], and (in)homogeneous broadening effects. This is physically logical, since any polaritonic (matter-like) mode increases the loss of the cavity. Thus, if a nanometer small source is desired, high optical loss is unavoidable. Lastly, any on-chip light source should be electrically driven; this requires electrical contacts, which hinder integration density if any photonic mode is used due to the avoidance of optical losses. Taken together, the physical mechanism of using free-carrier recombination across a semiconductor band gap for light emission bears fundamental drawbacks limiting on-chip sources. As such the demand for a silicon-based, nanometer small, potentially fast-modulatable, electrically-driven on-chip light source operating at room temperature remains unmet to date.

A possible option around such source-related bottlenecks is to turn to a different light creation mechanism; let us consider quantum tunneling of an electron across an electrically biased barrier (Fig. 1a,b). Here the electron can either tunnel elastically loosing energy to phonon modes, or inelastically creating a photon. The probability of the latter has been predicted to reach 10% for stimulated processes [19]. This mechanism is interesting for light sources for two reasons; a) since the temporal response upper limit of an tunnel event is governed by Heisenberg's uncertainty principle, the large optical energies of visible and NIR photonics demand sub ps-fast response times, and b) if the final goal is to create a laser, the requirement for carrier population inversion of the gain material could be simply met by biasing two Fermi seas against a doped semiconductor; in a band diagram one then has $10^{24}$ cm$^{-3}$ carriers from the metal residing above the $<10^{15-21}$ cm$^{-3}$ carriers of the semiconductor Fermi level. If one is able to convert these $10^3$-$10^9$ excess carriers into photons, a light source can be realized.

Indeed, biasing two Fermi levels against one another across a thin tunnel oxide (metal-insulator-metal (MIM) shows faint light emission [20]. The emission observed through the 10-100's nm thick top metal is expected to be miniscule, and is hence a poor indicator of the true internal photon conversion efficiency of the tunneling process. While lowering the device to cryogenic temperatures [21] or reducing the metal thickness in an ad-hoc manner shows small emission improvements. Thus, the outstanding challenge is to understand the underlying light-creation mechanisms based on inelastically-tunneling electrons. Moreover, the spectral observations from [22] are inconsistent to other reports, and the originally anticipated quantum condition ($E_{emission} = hv = qV_{bias}$ (1) [23], where $v$ is the frequency, $q$ the charge, $V_{bias}$ the applied voltage bias, and $h$ Planks constant) is experimentally not validated [20,22,24,25]. The argument of a convolution between the spectral-dependent tunnel current and the device-internal plasmon was made to explain the deviation from (1), as further validated in results presented here [26].

Here we report on the demonstration of an electrically-driven, CMOS compatible, silicon-based plasmon on-chip source based on inelastic electron tunneling operating at room temperature (Fig. 1). The source consists of a metal-insulator-(doped) semiconductor (MIS) tunnel junction of an silicon-on-insulator (SOI) substrate, which supports a sub-wavelength plasmonic eigenmode [27]. We show that the in free-space detected photons originate from plasmons generated inside the source, which are converted into photons via momentum matching facilitated by surface roughness or a grating. Accessing the dense plasmonic modes hidden underneath the metal layer via selective etching reveals high-field densities inside the device. We find an optimum top contact thickness close to that of the skin depth of the junction metal [28]. Matching the emission to free space via a sub-wavelength grating, we demonstrate a vertical surface-emitting source with a 40 fold enhancement outcoupling relative to flat metal film. Using the thermionic emission current with an added tunnel-barrier thickness-dependent current shows the possibility for 10's GHz-fast direct modulation for sub 1-nanometer thin tunnel oxides. Given our measured electroluminescence (EL) intensities, we estimate the wall-plug efficiency to be about half a percentage (see supplementary material) [29].

**Results**

The underlying physical processes of this source are quantum mechanical elastic- and inelastic electron tunneling across a barrier (Fig. 1a,b). In the elastic tunneling picture, a conduction electron in doped silicon approaches a biased, thin oxide barrier. With each scattering event at this barrier, there exists a finite probability that an electron will tunnel into the conduction band of the metal counter-electrode, conserving electron kinetic energy. After tunneling, the hot electron achieves thermal equilibrium with the electron sea via heat dissipation (Fig. 1a). Inelastic electron tunneling, in contrast, does not conserve electron energy; instead, a photon or phonon is created as the electron transfers to a lower energy state in the metal counter-electrode (Fig. 1b). The dominance of elastic tunneling over inelastic tunneling determines the conversion efficiency. The latter was predicted to occur with a probability of up to 10% when optimized for materials and bias [23]. While this efficiency needs to be yet validated, its probability is proportional to the photon density, which can be made dominant by introducing a cavity. Note, that the bound mode of this MIS system is sub-wavelength, thus allowing for a nanoscale cavity enabling device scalability. However, the cavity impact is not the aim of this work here, which focuses on demonstrating a silicon-based tunnel junction, experimentally proving its light creating mechanism while demonstrating device functionality.

The light source is comprised of a p-silicon substrate with a nanometer thin native oxide topped of with noble metal electrode on a SOI substrate (Fig. 1c). Such a metal/low-dielectric/high-dielectric structure supports a hybridized mode between a classical photonic waveguide and a surface plasmon polariton when placed on a low-index substrate (Fig. 1d) [5, 30, 31]. This geometry acting as an optical capacitor is able to confine light below the diffraction limit of light down to $(\lambda/20)^2$ at visible and NIR frequencies [30], while exhibiting a high electric field density inside the oxide gap. This field density has an impact at the time response of the device as discussed below. This mode has demonstrated Purcell Factors in the ten's to approaching hundred [32] making it a candidate for quantum electrodynamics studies. Applying a bias at the junction shows a surface emission that can be observed with the naked eye, or captured with a CCD camera (Fig. 1e). This emission is a plasmon-to-photon converted output assisted by the top-metal providing wavevector matching to free space, as discussed below.

Next, we show and discuss the observed electroluminescence when biasing the junction. Relating the emission to band-diagram under bias we explain the internal physics and operation principle of the source. The surface emission from the top metal (without a grating) is relatively weak, and shows randomly distributed hot-spots originating from localized surface plasmons driven by random current fluctuations [9]. The elastically tunneling electrons found spatially overlapping with the high-field density of the high k-vector hybrid-plasmon mode. Thus, the only optical mode to be excited is the hybrid plasmon polariton mode. These photon-plasmons leak through the top metal experiencing losses. Upon reaching the sample surface (metal-air interface), they form a surface plasmon polariton. The latter has high impedance to free-space, however, the grain boundaries of the electron-beam evaporated metal film ($RMS_{Au}$= 5-10 nm) add momentum to the plasmons, and thus scatter into free space. As such the external conversion efficiency of these MIS tunnel junction sources is understandably low and observed to be $10^{-5}$ or less [22, 23]. While improvements in the outcoupling efficiency are an engineering challenge, we first turn our attention to analyzing and validating the light creation mechanism.

The source' internal mechanism can be explained by a combination of the band diagram, tunnel current, and subsequent conversion into plasmonic modes under different biasing conditions (Fig. 2a). Electrically the junction is equivalent to a capacitor with a parallel tunnel resistor. The capacitor formed by the MIS stack has the three-known operation regimes from which accumulation (negative bias) and inversion (positive bias) are of importance for tunneling. Under forward bias to the metal, the semiconductor bands bend downwards (inversion) facilitating gate tunneling leakage current ($I_2$) from the metal to p-type silicon (Fig. 2a). Changing the bias polarity results in upwards band-bending; here the accumulated holes tunnel across the electrostatically-thinned oxide either elastic- and inelastically ($I_1$, Fig. 2a). This asymmetry is key leading to an accumulation-current magnitude that is significantly (~$10^9$) relative to the gate tunneling leakage current $I_2$ under inversion [supplementary information]. This explains our observed emission from the junction under negative bias only (Fig. 2b). With applied bias voltage the electric field and $I_1$ rises, creating hot electrons leading to emission events that feed two plasmonic mode asymmetrically [33]. Light emission originating from $I_1$ feeds both the hybrid-plasmon mode (via inelastic current tunneling) and the surface plasmon (via elastically tunneling hot electrons) (Fig. 2a) [20, 34]. The hybrid plasmon mode inside the junction is hidden from sight unless accessed otherwise (i.e. scattered out) [30]. These hybrid plasmons propagate through the top metal from where they scatter into free space via momentum added by the metal roughness.

Verifying the origin of the hypothesized photon emission process to the tunnel current, we obtain the current-voltage (I-V) characteristics and relate it to the integrated EL (Fig. 2b). Fitting the I-V curve to a tunnel current model gives a diode ideal factor of 1.35 corresponding to a tunnel oxide thickness, $t_{ox}$, of 2.6 ± 0.4 nm, which is within the expected range of the oxide used (i.e. native silicon oxide) [35]. Our results show that the I-V characteristic matches Fowler-Nordheim tunneling models [34, supplementary material]. The DC tunneling current responsible from light emission increases with applied bias voltage ($V_{bias}$) due to a higher electric field, which elevates the field densities of the two plasmonic modes by way of creation of hot electrons and thus the observed electroluminescence intensity (Fig. 2b). We find that both the current (measurement and simulation) and corresponding EL intensity track each other well, thus verifying that the tunnel current is indeed the origin for the plasmon creation. Note, each experimental data point presented in Figure 2b is an average across 15 different devices in order to account for sample variations. While the electric field across the tunnel oxide is high, the breakdown voltage for $SiO_2$ is ~3 V/nm [36]. This matches our experiments in which devices above 8 volts are mostly shorted. This also leads to a degraded light emission enhancement factor when comparing the junction EL from a flat top metal with that when a grating was etched as discussed below in Figure 3e.

The EL spectrum grows with voltage bias as expected, and shows two broad double-peak lineshapes centered around 720 nm and 550 nm (Fig. 2c). The overall shape is not simply given by the quantum condition $E = h\nu = qV_{bias}$, but depends on the convolution of the spectrally dispersive tunnel current density with that of the eigenmode of the system, namely the hybrid plasmon mode [26]. The spectral power density depends on current-fluctuations leading to plasmon creation; small fluctuations in the tunnel current lead to electric field fluctuations, which in turn accelerate and decelerated electrical carriers acting as a plasmon source. As such the spectral dispersion of the internal hybrid plasmon mode is material sensitive [26]. Following this line of thought, our modeling confirms the experimentally observed spectral double peak with exponential decaying emission intensity (Fig. 2c, see methods) [22, 26].

The plasmon generation takes place inside the junction feeding the junction's eigenmode. Thus, we explore the exact location of the plasmon creation, which we hypothesize to be either inside the tunnel oxide, or near the tunnel oxide inside the metal by physically accessing the mode via selective etching into the structure using focused ion beam milling (Fig. 3) [30]. In addition, we aim to increase the outcoupling efficiency, by thinning- down the emission-blocking top metal layer, and further introduce a grating design to add momentum to facilitate plasmon-to-free-space coupling discussed below. While MIM tunnel junctions are fundamentally limited in terms of out-coupling efficiency by the metal, our MIS junction can, in general, be flipped up-side-down to become a SIM and hence allowing for high normal surface emission. However, with our aim of using this source as a CMOS-compatible source in SOI platforms such as coupling emission vertically to waveguides or across interposers, we keep the underlying silicon layer at the bottom for technological relevance and ease of integration [37]. Other applications, however may require surface emitting sources such as lab-on-chip [38] and DNA sequencing or agent detection [39]. Considering a skin depth of about 30 nm for gold at visible frequencies, only 1.8% of the EL reach the top of the device for metal thicknesses of ~100 nm. On the other hand, if the metal is too thin, it resistive losses incur voltage drops, which diminishes the EL per applied voltage. The question is whether the thinning the top metal results in higher emission brightness. Gradually thinning the metal by a depth $d$, we find a linearly exponential EL increase matching the Beer-Lamber law (see Methods, and region A, Fig. 3a,b). Until this point we thinned the metal on the entire device pad area and utilized the aforementioned momentum matching from the metal roughness [9, 22, 23]. In the limit of etching (removing) the entire pad, light creation would reduce to zero as no tunnel junction is formed. However, if we etch only certain regions passing through the light-creating oxide layer, the resulting sharp edges facilitated the EL to scatter to free space by providing high wavevectors. On the other hand, one loses light creation at these areas that were etched into the silicon (Region B, Fig. 3b). Thus, with outcoupling intensity in mind, optimization becomes a function of grating fill-factor; an initial test for a 1:1 fill factor shows a sharp drop for the case of etching deeply into the silicon (Region B, Fig. 3b). This is due to the lost photon creation and possible parasitic losses introduced by the Gallium beam doping from the focused-ion-beam (FIB) milling process used, explaining the roll-off (d > 150 nm, Fig. 3b). Another reason for the weak light emission intensity is attributed to the fact, that both the hybrid photon plasmon and the surface plasmon modes are intrinsically nonradiative [40], because a) their relative wavevector is approximately one order of magnitude different, and b) both mode's wavevectors are larger (at least 2x) compared to that of free-space [41]. However, the maximum output enhancement (33x) of the etched regions versus the metal pad no grating (MPNG), is observed when the metal is not entirely removed, but about 20-30 nm of metal remains (Fig. 3a,b).

Next, we investigate the outcoupling efficiency of a grating and compare it with that of the thinned metal film (Fig. 3c). In order to improve the out-coupling ratio and gaining further inside into the Silicon-SPP light source, we etched selective gratings into the device accessing the hybrid-plasmon-mode and facilitating coupling into free space (Fig. 3d,e). For instance, the EL originating from a single groove (Fig. 3c, Area$_{1x\text{-groove}}$ = 0.05 μm$^2$) is about one order of magnitude brighter than the total integrated power of the entire pad (49 μm$^2$). However, accounting for all inefficiencies we estimate the total external quantum efficiency of these tunnel junctions including the etched grooves to be on the orders of 0.01 - 0.1%; from the 4π emission sphere only a fraction of the upper half dome is captured by the objective lens (NA = 0.42) and converted into an observable by the camera (QE~ 75%), and the hybrid-plasmon mode-to-free-space coupling efficiency is about one percent (Fig. 3e). However, the light emission can be observed with the naked eye (Fig. 1e), and an example video was captured by a CMOS camera (supplementary online material). Furthermore, the tunnel probability into the light emission mode can be enhanced via the Purcell effect, reducing the tunnel current resistance. With experimental Purcell enhancements of about 100 being demonstrated [32], one could expect the conversion efficiencies to approach single digit percent range, which is just about 10x away from the predicted values [19]. Next we compare the emission between free space and the plasmonic mode with a designed grating [40, 42] (supplementary material, Fig. 3d-f). We find that a square-shaped grating does not offer a significant improved outcoupling efficiency over a thin (20 nm) thick metal layer, because the island-like surface roughness of the poly-crystalline deposited metal film (with an optimum thickness) already provides some momentum to facilitate outcoupling (Fig. 3e) [26,43]. Optimizing the grating up to about 40x times enhancement in light emission intensity was observed via the sine-shaped grating in comparison to the square shaped grating (Fig. 3e). These results can be explained from a more homogeneously provided momentum added to the hybrid plasmon mode. Furthermore, we find that the grating efficiency improvement over the non-grating case grows with bias and drops after ~5V which is likely due to joule heating and consequently thermal stress on the junction (see supplementary material, Fig. 3e). This indeed is consistent with our experimental observations, where the device lifetime is inversely proportional to the bias voltage. An optimum operation voltage is at around 3-4 volts for optimized light emission while keeping the thermal stress low (Fig. 3e).

Because the emission originates from the rapidly-thermalized Fermi-sea of a semiconductor and into the conduction band of a metal, the limiting processes do not follow the standard rate equations of light emitters and lasers. From Heisenberg's uncertainty principle, large optical bandwidths imply inelastic tunneling speeds on the order of 10's of femtoseconds. Indeed, the temporal response of a tunnel event has been measured as short as 100 atto seconds [44]. Comparing this to recombination lifetimes of direct and indirect bandgap semiconductors such as GaAs and Si, which are on the order of nanoseconds and milliseconds, respectively, we find that tunnel junctions may allow for a high modulation speed [45, 46]. With the delay of the actual tunnel being negligible, we analyze the electrical circuit-related constrains to understand the junction's actual response time. The limiting factor is related to resistive and capacitive (RC) effects of the junction itself [47]. Our MIS tunnel source is a planar structure acting as a parallel plate capacitor (Fig. 4). Here, the relevant resistance in this case is not the line impedance but the resistance. For inelastic tunneling events to be dominant, the electron tunnel current must dominate the displacement current across the capacitor. Note, that the tunnel resistance scales inversely with area, whereas capacitance scales linearly, yielding an RC time constant invariant to area in an ideal case (i.e. neglecting nonlinearity, see methods). However, the tunnel resistance scales exponentially with thickness, while the capacitance scales only linearly. As a result, high modulation speed (>40 GHz) can be achieved with sufficiently thin tunnel oxides (0.6 nm, Fig. 4). The latter is technologically achievable by adjusting the deposition cycle in the atomic-layer-deposition process (Fig. 4) [48, 49]. A video of a dynamically modulated junction that is slow-enough for the millisecond response time of digital camera is provided in the supplementary online information. Replacing the top metal of the MIS junction with a poly-silicon drops the speed to ~8 GHz due to lower tunneling current originating from the higher resistance of the doped semiconductor vs. the metal (Fig. 4). It is worth mentioning, that the metal serves a triple function in this light source; (i) metal confines the optical mode enabling device scalability via allowing for sub-diffraction limited modes as we have previously shown for hybrid plasmons [29], (ii) metal is a heat sync, since replacing the top metal with poly Silicon raises the device temperature enabling higher modulation speeds (i.e. $P_{dissipated}$ = $E$/*bit* x *bitrate*) (see supplementary information) [29], and (iii) metal acts as an electrical contact allowing for low-voltage drops in the contacts leading up to the device. The latter is not possible for photonic devices as their optical loss from heavy-doped (low resistance) semiconductors is detrimental to the insertion loss devices [49]. Interestingly, the modulation speed can be accelerated further by increasing the inelastic tunneling probability via the Purcell factor, which could be achieved by introducing nanoscale cavities [32,50,51]. Such acceleration of emission processes will thus further decrease the tunneling resistance, hence increasing direct modulation speed. This also leads to an enhanced quantum efficiency and thus wall-plug efficiency, in analogy to the spontaneous emission factor reducing the laser threshold [19].

In conclusion, we experimentally demonstrated an electrical-driven Silicon-based plasmon source where the plasmon creating mechanisms originates from inelastically tunneling electrons across a quantum mechanical thin dielectric barrier. This light emitting tunnel junction is not bound by physics of a classical semiconductor 2-level system, but by the probability of plasmon creation from inelastically tunneling electrons. We validate the origin of the plasmon source to the tunnel current, and explain the spectrum with dispersive tunnel-current probability modeling matching our experimental results. We access the high-density plasmon mode inside the tunnel junction by slicing the device open, and find the optimum light emission condition to be an interplay between the internal and surface-bound optical modes, propagation and coupling losses, and wavevector matching. We find the ideal metal thickness of this metal-insulator-semiconductor junction to coincide with the skin depth of the metal used. However, introducing a grating increases the outcoupling efficiency to free space 40 fold relative to the optimized metal thickness. Lastly, the broadband emission of the tunnel events suggest a fast modulation speed limited given Heisenbergs uncertainly principle. The actual modulation speed for devices scales inversely exponentially with the tunnel-barrier thickness and thus with the tunnel current. Our results show a viable path for silicon-based light emitters based on technological-relevant Silicon-on-insulator platforms, suitable for electrically-driven sources operating at room temperature in optical and hybrid plasmonic networks on-chip [52,53].

**Acknowledgements**
V.S. is supported by Air Force Office of Scientific Research under the Award FA9550-17-1-0377. We thank Josh Conway for helpful discussions, and Ergun Simsek for numerical analysis support.

**Author Contributions**
V.S. conceived the idea. H.G and V.S. performed experiments and generated figures. H.G. performed both analytical and numerical analysis. FS performed analytical analysis. All authors discussed the results, and wrote the manuscript.

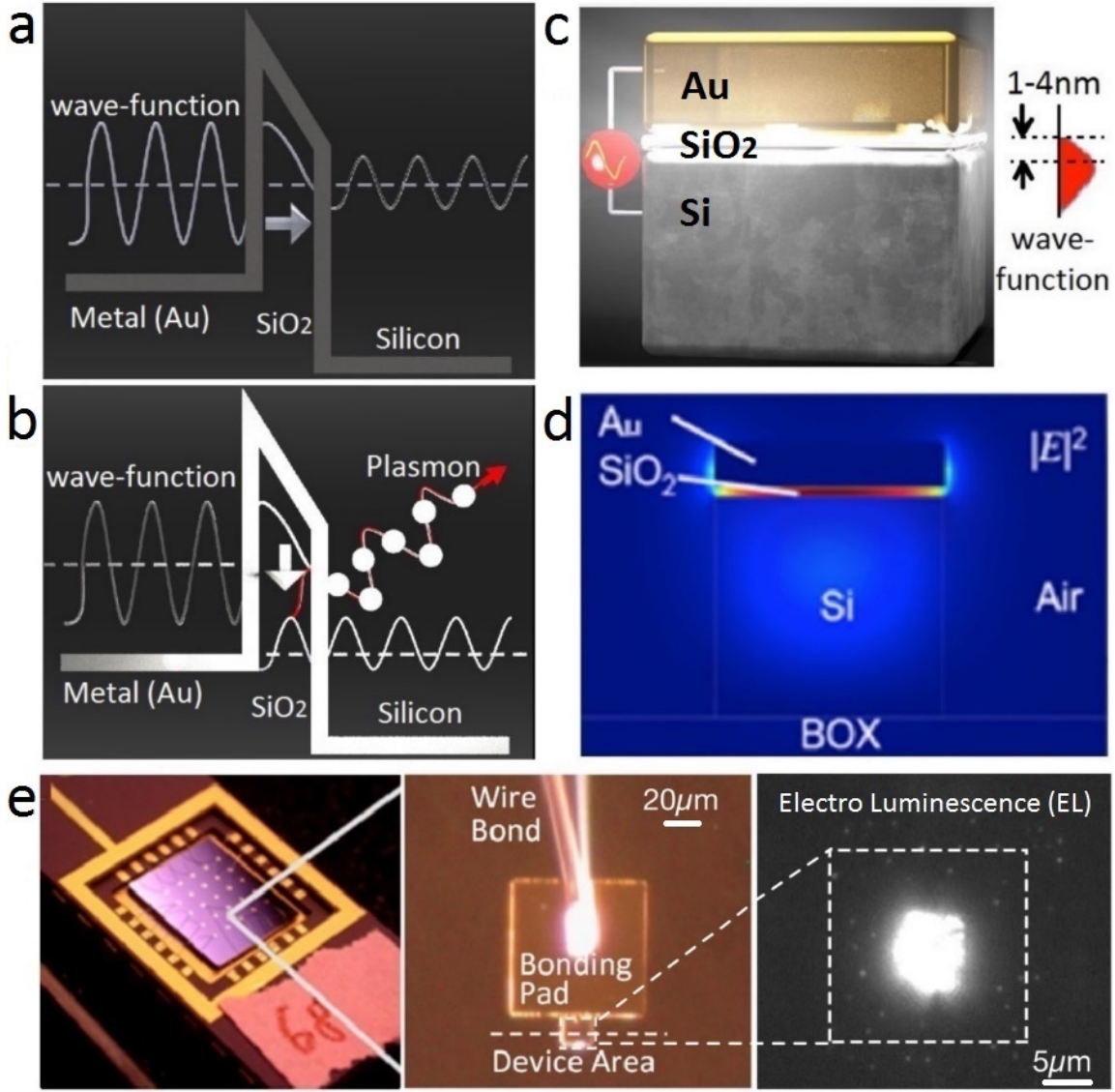

**Figure 1. Silicon plasmon light source operation mechanisms and device details.** Schematics of (**a**) elastically- and (**b**) inelastically scattering carrier tunneling, the latter leading to plasmon creation. (**c**) Electrical bias scheme for the light-emitting tunnel junction based on a metal insulator semiconductor configuration. Metal: Au = 100 nm, Oxide: $SiO_2 = t_{ox} = 2.6 \pm 0.4$ nm, Silicon: thickness = 200 nm, p-type on a silicon-on-insulator (SOI) platform. Side-lengths of the squared devices vary from 0.5-40 μm. (**d**) Subsequent sub-wavelength plasmon hybrid mode inside the junction. Highest |$E_y$|-field strength is inside the thin tunnel oxide [5, 30, 31]. For surface normal emission, field density of the hybrid plasmons leak through the metal reaching the metal-air interface forming a surface plasmon polariton. When momentum is added, plasmons couple into free space (hot spots in electroluminescence in e). (**e**) Tunnel junction devices on chip carrier are wire bonded for electrical contact with spontaneous emission observed from the tunnel junction area. A grating facilitates scattering into free space, which is visible with the naked eye.

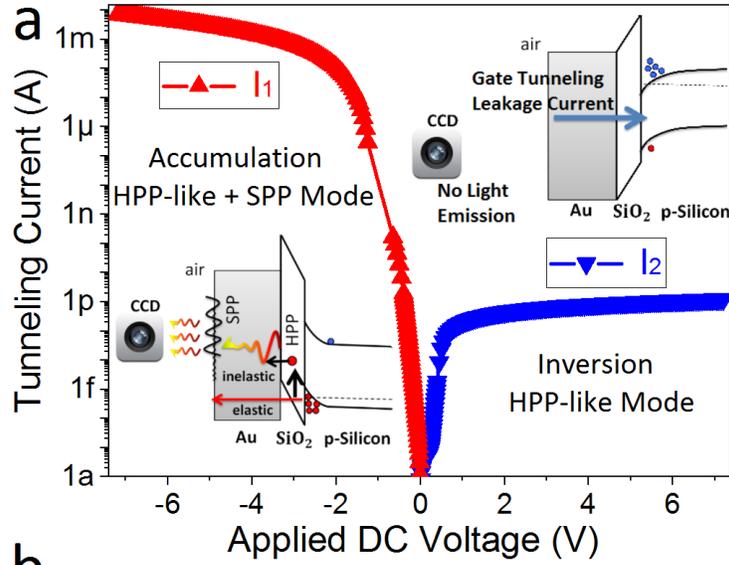

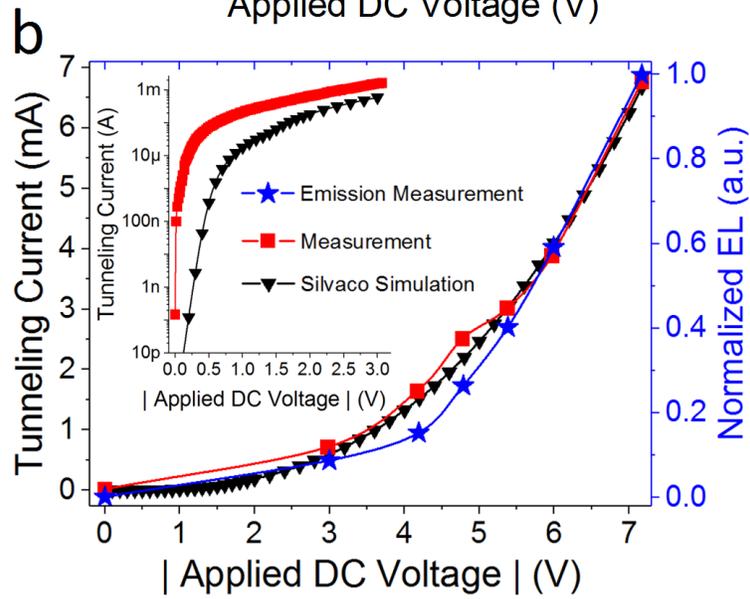

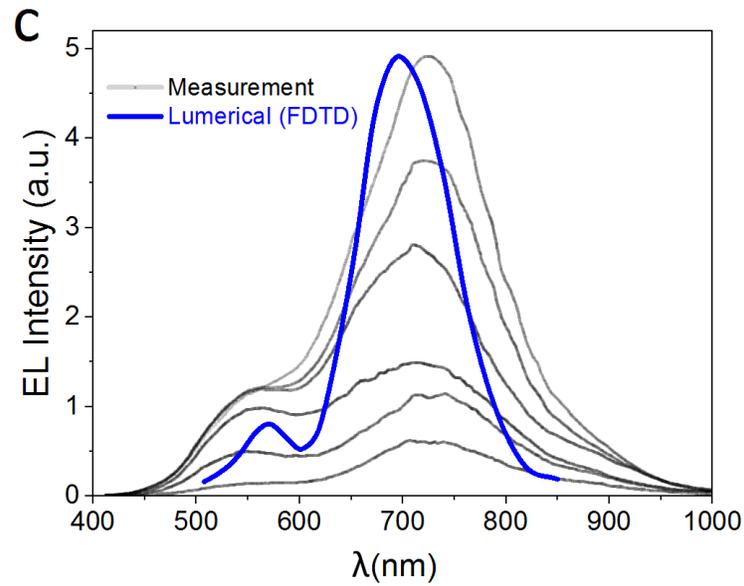

**Figure 2. Plasmon and light creation mechanism of the tunnel source.** (**a**) I-V curve for the device and current relation with respect to optical modes. The light source is a MIS capacitor resulting in higher emission when biased in accumulation ($V_{bias} < 0$) as opposed to inversion ($V_{bias} > 0$), due to the corresponding tunneling current feeding two plasmonic modes. Materials: p-type Si (0.1-2.8 Ωcm), SiO$_2$, and Gold. Numerical tools: Silvaco. Insets; band diagrams and mode creation; a hybrid-photon-plasmon (HPP) like mode, and a surface plasmon polariton (SPP) are fed by the accumulation tunneling currents. (**b**) I-$V_{bias}$ characteristic shows a bi-exponential growth. Each data point is an average of 15 devices verifying that the tunnel current is the origin for the plasmon creation. Fitting the current to a tunnel model yields an ideal-factor of 1.35 corresponding to an oxide thickness, $t_{ox}$ = 2.6 ± 0.4 nm. The light output was found to increase superlinearly with current. A 7 μm x 7 μm plasmon emitting tunnel junction in Silvaco was modeled with 2.1 nm thick SiO$_2$ (effective mass of 0.2m$_o$ for electron and 0.045m$_o$ for holes) with a p-type doping concentration of $2\times10^{17}$cm$^{-3}$ for silicon. (**c**) The junction's spectrum is centered around 720 nm increases with bias (1-6 Volts). The spectral lineshape is not simply given by the quantum condition $E = hv = qV_{bias}$, but depends on the convolution of the spectral tunnel current density with that of the hybrid plasmon polariton mode [26]. Lumerical FDTD simulation's result for 20 nm thick Gold (with surface roughness), when dipoles are used as a light emission source match the measurement results.

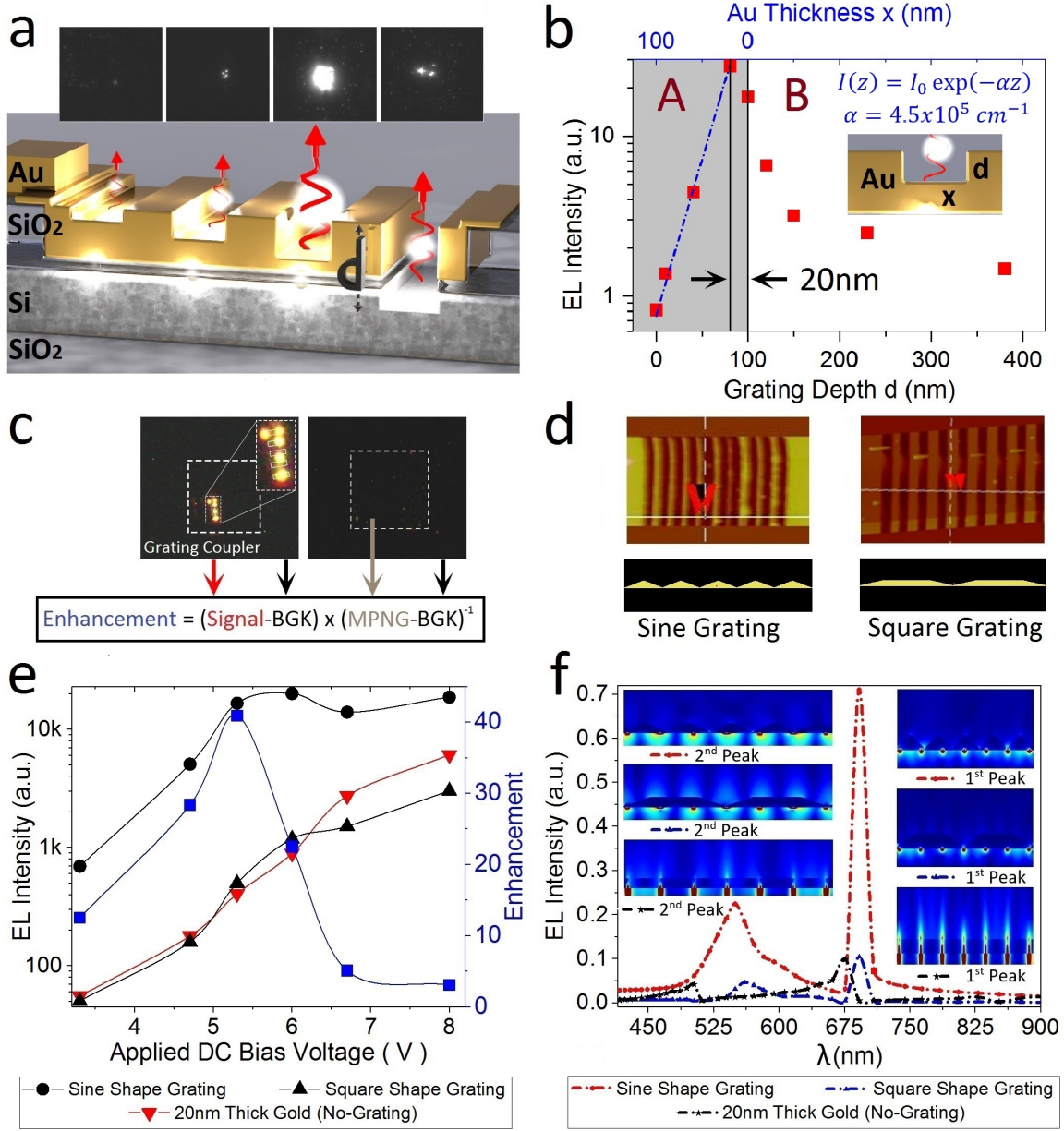

**Figure 3. Light emission to free-space coupling.** (**a**) Schematic of etched device area using focused-ion-beam (FIB) milling to reduce the blocking top metal (Au, thickness = 100 nm). $d$ = etch depth. Inset: far-field CMOS images of the electroluminescence (EL) showing an optimum thickness. (**b**) Collected EL as a function of etched device pads (normalized for area, red squares). The exponential increase is due to the trivial loss reduction when the metal is thinned down. Grey area represents metal thickness. Optimum outcoupling thickness is about skin depth of metal (~20 nm). Absorption coefficients: $(\kappa_{Au})_{fit} = 5 \times 10^5$ decently close to $(\kappa_{Au})_{[54]} = 6 \times 10^5$. (**c**) Formula and indication for enhancement factor calculation. S = signal, BGK = background, MPNG = metal-pad-no-grating. (**d**) Sine shape and square shape gratings view under AFM (**e**) The EL enhancement for a sinusoidal grating yields an enhancement of up to 40 times. The degradation of the EL curve for high bias voltages is due to both imperfect MIS-junction and thermal stress from joule heating. (**f**) EL performance comparison of Sine shape, square shape and 20 nm thick no-grating plasmon emitting tunnel junction in Lumerical FDTD simulation.

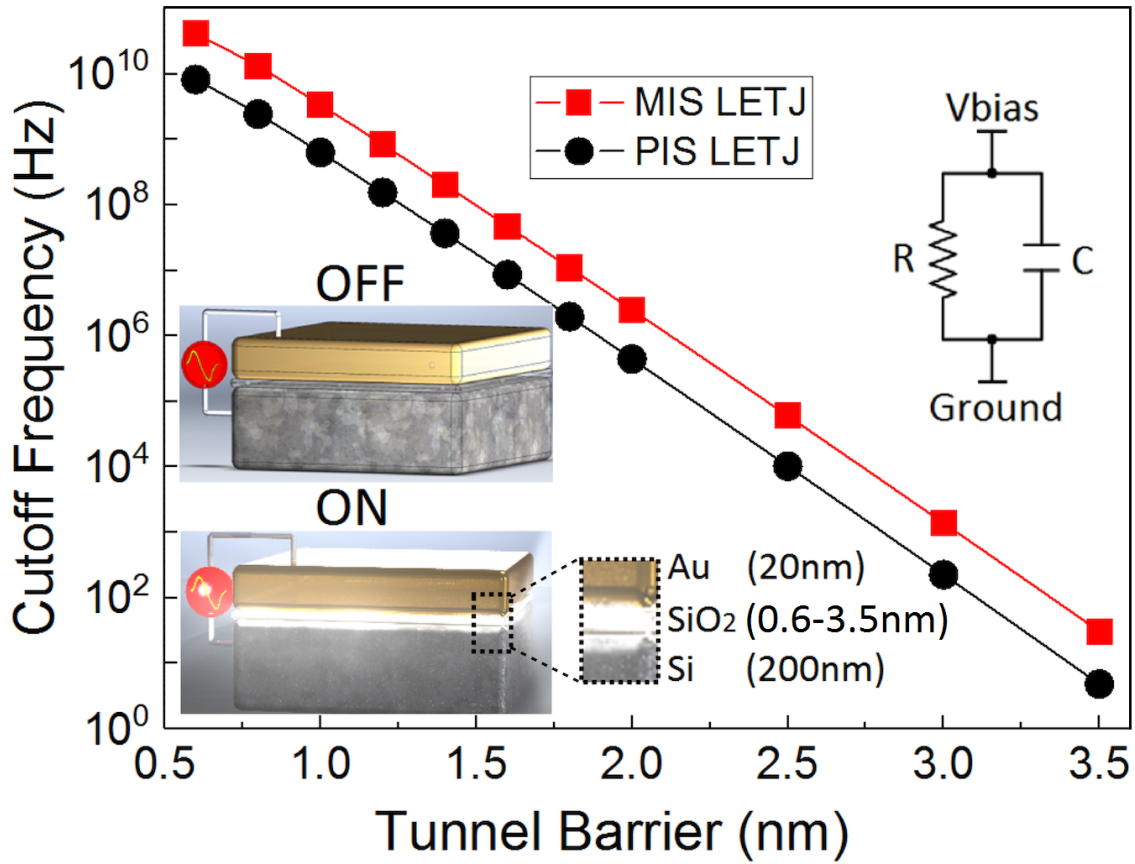

**Figure 4. Direct modulation speed of the tunnel junction.** Results show that at 10's of GHz-fast modulation are possible for tunnel oxides less than one nanometer ($V_{bias}$ = -3.4V). MIS = metal-insulator-semiconductor and PIS = polysilicon-insulator-semiconductor. Silicon thickness = 200 nm Inset: plasmon emitting tunnel junction layout and its equivalent circuit model.

## Methods

**1) Experimental Methods:** The fabrication process starts with piranha cleaning and buffer oxide etching (BOE) to clean the native oxide layer on top of the Silicon substrate. Next, 300 nm thick SiO$_2$ was grown on Silicon via PCVD. Following, photolithography and BOE (70 nm/s) was conducted together to pattern and remove the SiO$_2$ from the device areas. Next, 2-3 nm tunnel oxide was grown via O$_2$ flow for 1.5 hours at 700 °C. Next, photolithography and lift-off processes were conducted together to create 100 nm thick top metal (Au), and a final metallization step to create 120 nm thick contact pads (Cr/Au). IV measurement: DC probes located on the device's pad are connected to a semiconductor parametric analyzer. DC voltage increased with small step up to 5-6 V and data recorded to plot the IV response. The measurement was repeated with opposite voltage polarity applied to the device to plot both accumulation and inversion region. Luminescence measurement: Sample was located on probe station with optical probes aligned towards light emission direction after proper calibrations. The light emission was collected and sent to an OSA (optical spectrum analyzer) to observed emission with respect to wavelength at room temperature. Light emission was pictured and recorded via a CMOS camera.

**2) Numerical Methods**

*1. Tunneling current as a function of applied bias voltage:* A 7 µm x 7 µm square shape MIS structure in this work (Fig. 2c) was rebuilt in the Silvaco with 200 nm thick silicon and additional

212 $\mu$m thick silicon to replicate the resistance coming from wires, following by 2.1 nm silicon dioxide (SiO$_2$) and gold (Au) layer. A uniform p-type doping with a concentration of $2x10^{17}$ was applied to the Silicon while defining effective mass of 0.2m$_o$ for electron and 0.045m$_o$ for holes in SiO$_2$ layer. The carrier lifetime for both holes and electrons in silicon was introduced as $10^{-6}$. The numerical solver used for electron, hole and band-to-band tunneling mechanism (self-consistent quantum tunneling model) while using additional convergence tools inside Silvaco to improve the accuracy. In the same way, the tunneling current to calculate the cutoff frequency of the MIS device (Fig. 4) was derived from Silvaco where the Silicon thickness is 200 nm for both PIS and MIS device.

*2. Emission spectrum and grating dependent emission intensity:* MIS structure (Fig. 1c) was rebuilt in Lumerical FDTD with the dipoles located in oxide region to replicate the light emission. A surface roughness was created on 20 nm thick Gold surface for the spectrum. Gold data from [54]. For grating the study; the dimensions of the fabricated grating structures were measured with AFM and the data were used to reproduce the same structures in Lumerical FDTD. The oxide and the silicon were removed from the structure and only 180 nm thick metal grating structure on top of 20 nm thick metal plate were kept. Dipoles were located under the 20 nm thick plate to imitate the light emission. The structure has a specific FIB beam angle between gratings and only the grating duty cycle was varied during the simulation (see supplementary material). Additional simulation was conducted after removing the grating structure to obtain the wavelength response independent form the grating effect for 20 nm thick Au layers. A high mesh FDTD (8/8) with a stability factor of 0.95. An extra high mesh box was added to get further improvement in accuracy. A field time monitor was used to properly define the simulation time. A mode expansion monitor was added to resolve the effective index and mode. A frequency domain profile was used to obtain the light emission profile. A frequency domain field and power monitors were used to measure both reflection and transmission. The light emission intensity for different duty cycle was recorded by using data from transmission and reflection monitors.

### 3) Analytical Methods

*1. Emission Loss and Grating Analysis:* Thinning the metal, the EL increases exponentially according to Beer-Lamber-Law; $I(z) = I_o \exp(-\alpha z)$, where $I$ is the emission intensity, $I_o$ is the initial amplitude of the emission intensity, z is the metal thickness towards the propagation direction and $\alpha$ is the absorption coefficient of the metal. The light emission intensity ($I_e$) with respect to grating parameters ($\Lambda, d, \theta$, and $\epsilon$) [26];

$$I_e \approx \frac{P(k,w)}{\exp\left(\frac{\lambda_e}{L_x}\right)} \quad \text{where } L_x = \frac{1}{2k_i}, k = K\sin\theta \pm nG, \ G = \frac{2\pi}{\Lambda}, \ k, k_i \sim \frac{\epsilon}{d}$$

where w is the corresponding wavelength, k is the momentum, n=1,2,3..., $\theta$ is the emission angle, $\lambda_e$ is the effective wavelength, $\Lambda$ is the grating period, $d$ and $\epsilon$ are the thickness and dielectric constant of oxide layer respectively.

*2. Tunneling Current and Direct Source Modulation Speed:* The tunneling current responsible from the light emission is given in [34] and proportional to the Electric field in the oxide layer as shown below;

$$I \ \alpha \ E_i^2 \exp\left[-\frac{4\sqrt{2m^*}(q\phi_B)^{3/2}}{3q\hbar E_i}\right]$$

where $E_i$ is the electric field in the oxide region, $m^*$ is an effective mas, ℏ is plank constant, and $\phi_B$ is barrier height. The electric field in the oxide region is proportional to the surface potential;

$$E_{ox} = \left(V_{bias} + \phi_{Bs} - \phi_{Bm} + \frac{k_BT}{e}\ln\left(\frac{N_C}{N_D}\right) - \psi_s\right)/t_{ox}$$

where $V_{bias}$ is the applied bias voltage, $\phi_{Bs}$ is the difference between the electron affinities of Si and SiO$_2$, $\phi_{Bm}$ is the difference between the electron affinity of $SiO_2$ and the work function of the metal, $t_{ox}$ is the oxide thickness and $\psi_s$ is the surface potential. The equivalent circuit of the MIS diode was derived as parallel RC [55] and the cut off frequency of RC is;

$$f_c = \frac{1}{2\pi RC}, \quad R = \frac{V_{bias}}{JA}, \quad C = \epsilon_o \epsilon_r \frac{A}{t_{ox}}$$

where $J$ is the tunneling current density, and was derived from Silvaco tool.